\documentstyle[preprint,aps]{revtex}

\begin{document}
\draft
\title{Can Density-matrix renormalization group be Applied

to two dimensional systems}
\author{Shoudan Liang and Hanbin Pang}
\address{Department of Physics, Pennsylvania State University,
University Park, Pennsylvania 16802}
\maketitle
\begin{abstract}

In order to extend the
density-matrix renormalization-group (DMRG) method
to two-dimensional systems,
we formulate two alternative methods to prepare the initial states.
We find that
the number of states that is needed for accurate energy calculations

grows exponentially with the linear system size.
We also analyze how the states kept in the DMRG method manage
to preserve both the intrablock and interblock Hamiltonians,

which is the

key to the high accuracy of the method.
We also prove that the energy calculated on a finite cluster
is always a variational upper bound.
\end{abstract}

\pacs{05.30.-d,75.10.Jm,02.70.+d}
\newpage
\narrowtext

Exact diagonalization is often the only reliable technique

for obtaining the ground-state
properties of interacting many-body systems.  Since the

Hilbert space
grows exponentially with the size of the system,
current computing power limits the diagonalization to

systems too small to infer the thermodynamic limit for

most problems.

One method that can potentially overcome this difficulty

is the
renormalization-group technique, in which one attempts

to describe

the many-body ground state with fewer degrees of freedom.

Unfortunately, this method, which works so well for the Kondo
problem$^{1}$ gives
 rather inaccurate results when applied to lattice models of

strongly correlated systems$^{2,3}$.

In an exciting recent development, the key weakness of

the previous

method has been identified$^{4}$ and a density-matrix

renormalization-group
(DMRG) method has been developed by White$^{4,5}$. When

applied to one-dimensional spin chains,

the DMRG method is highly accurate.

This success arises the hope that the unconventional ideas$^{6}$

proposed for the two-dimensional Hubbard model can now be tested,
allowing progress to be made in the theory of

high-$T_c$ superconductors.
In this paper, we take a step in this direction by devising ways to

perform two-dimensional
calculations and study the characteristics of the DMRG method when
applied to two-dimensional problems.
We find that the real-space DMRG method has an intrinsic

difficulty in

two-dimensional problems.  The amount of computation needed grows

exponentially with the linear system dimension. To identify

the source of the difficulty, we develop a physical picture

which also provides insights as to why the DMRG method works better

than the conventional real-space

renormalization group.
In addition, we prove that the finite-cluster DMRG method

always computes variational upper bounds to the

ground-state energy.

To analyze the physics of the DMRG method, we use the

simplest Hamiltonian of free spinless fermions.
We find that, for an $L\times L$ lattice,
the error in the calculated ground-state energy is proportional
to $e^{-m/m^*}$, where $m$ is the number of internal

states kept in the

calculation.
The coefficient $m^*$ grows exponentially
with the linear system dimension: $m^*\propto \alpha^{L}$ with

$\alpha=3.9$.$^{7}$
However the accuracy of the method mainly depends on

the total strength
of the inter block interactions, and is almost independent of the

length of the boundary. Thus, the
DMRG method should work well in, for example,

one-dimensional systems with
long-range but fast-decaying interactions.

The DMRG method is a diagonalization technique
in which one attempts to use a small
number of states to expand the ground-state accurately.
One divides the lattice under study into two parts, {\it i.e.},
the system and the environment, and asks if one is allowed to
use only $m$ states each for the system and the
environment, which $m$ states one should choose?  The answer
given by the pioneering work$^{5}$ of
White is the following:
construct the density matrix of the system
from the ground-state wave function of the whole
lattice by summing over the Hilbert space of the environment
and pick the $m$ largest-weight
eigenvectors of the density matrix. This is different from the
conventional approach where, instead, the $m$ lowest-energy

eigenvectors of the system are chosen.
Since the ground-state wave function of the whole system
is not known beforehand,
one constructs an approximation to it.  The DMRG method

systematically improves the approximation.
In addition to the ground state, this
procedure also works for low-lying excited states.

In order to study the characteristics of the DMRG

for two-dimensional

systems, we consider a simple tight-binding

Hamiltonian of spinless fermions defined by:
\begin{equation}
H=-\sum_{\langle ij\rangle } t_{ij} c_i^{\dag} c_j + h.c.,
\label{HAM}
\end{equation}
where $c_i^{\dag}$ ($c_i$) is the

fermion creation (annihilation) operator;
the sum is over nearest neighbors and $t_{ij}$ is
equal to $1 (t')$ for the bonds indicated by the solid (dashed)

lines in Fig. 1.
When $t'=0$, the model describes nearest-neighbor
hopping of electrons in a one-dimensional chain,

whereas when $t'=1$

the two-dimensional model is recovered.
For technical reasons, we restrict ourself to the case of
free boundary conditions.

The two-dimensional system is thus mapped onto a

one-dimensional model
with long-range hopping (Fig. 1).
The sites are numbered sequentially by integers along

the one-dimensional

lattice (the solid lines in Fig. 1)
The partition of the lattice in Fig. 1 minimizes the

interaction between blocks when the system is divided
into four blocks
indicated in parentheses:
$(1,...,i-1), (i), (i+1), (i+2,...,N),$
where $N=L_x\times L_y$ is the total number of sites.
We will denote the blocks by $(1), (2), (3), (4)$.
The blocks $(2)$ and $(3)$ are each composed of

only a single site.

A block is represented by states in sectors
which are labeled by $n$, the number of fermions.
The Hamiltonian matrix in each
sector and the matrix elements of $c_i$ completely
specify the quantum state of the block.
Using this information, the ground state of the
whole lattice can be obtained in the Hilbert space of
the direct product of the states from the individual blocks.
With this ground-state wave function, the density matrix
of blocks $(1)$+$(2)$ is calculated.
After the density matrix is diagonalized, the $m$

largest-weight states
are retained to form a new block $(1)$. The number of sites
in block $(1)$ has thus grown by 1. A complete iteration

starts with block
$(1)$ consisting of one site and continues until it

includes $N-3$ sites.

An approximated block $(4)$ is first
used as the environment block in calculating the ground-state
wave function. After each iteration, the
newly calculated block $(1)$ is ``reversed'' along
the one-dimensional lattice, replacing the corresponding
block (4) to serve as a better approximation for

the environment block.
This self-consistent procedure converges quickly

after a few iterations.

The trickiest part of the calculation is the
preparation of the initial states,
{\it i.e.}, the environment blocks.
Two methods are used.  In the first method we
perform a one-dimensional calculation (by setting $t'=0$) using
the method of White$^{5}$. In the second
iteration we set $t'$ to 1 and use the matrix elements of $c_{i}$
generated from the one-dimensional iteration to calculate the

interactions between blocks (1) and (4).
We find that $m$, the number of states kept, for $t'=1$ has to be
larger than $m$ for $t'=0$ in order for the system to converge to
the two-dimensional wave function.  One drawback of this method is
that the matrix
elements of $c_{i}$ far away from the end point (where two
blocks meet) are not preserved well because
they are not needed in the one-dimensional calculation$^{10}$.

To overcome this difficulty, we have also devised a
second method in which we prepare a one-dimensional chain
with $L_x+1$ sites as a substitute
environment block. The chain is diagonalized

using the conventional
method. The lowest few excited states, as well as

the matrix elements
of $c_{i}$ between them, are calculated. The point is

that in the DMRG

method, the
states saved are such that the matrix elements
of the $c_{i}$ near the end point of the blocks
are kept much better than those away from the end point,
whereas the lowest-energy eigenvectors do not have such a bias.
However, in this method, the density of states of the environment
block is not correctly represented.

In both methods, we also find that it is sometimes
necessary to iterate a few times with $t'>1$

before setting $t'$ to 1.
This helps the system to become two dimensional.

We emphasize that

once the wave function converges to the

two-dimensional solution, the
final wave function depends only on $m$ and
is independent of which
initial states are used$^{8}$.

To obtain the wave function
for the whole lattice,
blocks are combined taking into account the
conservation of $n$. The Hamiltonian matrix is sparse and is
calculated once and used repeatedly. To
diagonalize the matrix, we use the Lanczos algorithm
with selected orthogonalization,$^{9}$
which avoids the ``ghost vector''
problem with minimum added cost.
It is especially useful when several eigenvectors are sought.
We have found this method to be reliable and convenient.

We studied the error $\Delta E$ in the total energy for the

free-spinless-fermion Hamiltonian [Eq. (\ref{HAM})

with $t'=1$] on an
$L\times L$ lattice at half filling ($n=L^2/2$) and with free
boundary conditions. We found that the error in total energy

decreases exponentially
with $m$, the number of internal states kept

for the block (see Fig. 2).
However, to keep the error within a predetermined value,
$m$ must grow exponentially with $L$: the data for
$L=3, 4, 5, 6$ and $7$ are well described by

\begin{equation}
\Delta E = C e^{-m/m^*}, \;\; m^* = \alpha^L/A,
\label{DECAY}
\end{equation}
 with $\alpha(t'=1)=3.9$,
$A(t'=1)=185$; $C$ depends weakly on $L$ and approaches 1 at

large $L$.$^{11}$
We have kept up to $100$ states for the most difficult case
of the $7\times7$ lattice
where the error is about $1\%$. Extrapolation to $m\rightarrow
\infty$ can always be used from the finite-$m$   data in order
to get better estimates of the ground-state energy.

This result suggests that it is extremely difficult to
apply the method to two-dimensional systems. However,
the number of states needed in the DMRG method grows much slower
with the lattice size than $2^{L^2}$ states required
for exact diagonalization. The DMRG method may therefore still be
the best among the existing numerical methods for strongly
correlated systems. The exponential growth is intrinsic
to the DMRG method and is independent of methods used to prepare
the the initial state as long as the wave function

converges to the

two-dimensional one.

We next examined the feasibility of applying the DMRG method to

quasi-two-dimensional systems. We studied how the error
depends on $m$ for stripes $L_x\times L_y$ with $L_y>L_x$.
Based on several calculations on $3\times 8,\;

3\times 10,\; 3 \times 12$
lattices, and wider stripes with $t'=0.5$, we found that

when $L_y\ge 2L_x$, $m^*$
becomes independent of the chain length $L_y$

and depends only on $L_x$.

We then studied the dependence of $m^*$ on $L_x$.
For $t'=1$ and $L_x\times 10$ stripes with $L_x = 2,3,4$,
we found that $\Delta E $ decreases exponentially
with $m$. The decay constant $m^*$ grows with $L_x$ following
$m^*\approx 4^{L_x}/4$.

An interesting question is whether $m^*$ depends

on the number of contact
points only or whether it also depends on the

total interaction strength

between blocks $(1)$ and $(4)$.
To answer this question, we varied the interblock
interaction strength $t'$.
We found that $m^*$ depends strongly on $t'$
and weakly on the number of ``contact points''.
For $t'=0.5$, we studied the $L\times L$ system with $L=4,5,6$.
Fitting to Eq. (\ref{DECAY}) we obtained $\alpha(t'=0.5) = 1.9$ and

$A(t'=0.5)=3.3$.
For the same lattice geometry, the number of

states required is much less
for $t'=0.5$ than for $t'=1$. In addition,

when we use stripe geometry
$L_x\times 10$ with $L_x=2,3,4,5$ we

obtain $m^*=(2.1)^{L_x}/2.2$.
The data are consistent with the idea that
the number of states required grows with the total
interaction strength between two blocks (which is proportional

to $t'L_x$): $\log _{10} (m^*)\propto t'L_x \log _{10} 4$.

The possibility that $m^*$ depends only

on the number of contact points
is easily understood because the success of the DMRG method tells
us that the boundary effects are very important.
The Hilbert space of the boundary sites should be kept well.
The total number of states needed grows

exponentially with the length of the
boundary. On the other hand, the

possibility that $m^*$ depends on the
interblock interaction strength is readily
understood in the conventional

renormalization-group picture.
If two blocks described by a few

lowest-energy eigenvectors

interact with matrix elements $t$,
we know from perturbation theory that the energy levels
of both blocks within $\delta E \approx t$ of the ground state
will be mixed in the ground state of the superblock$^{12}$.
Since the number of states within energy
$\delta E$ of the ground-state energy increases exponentially
with $\delta E$, the number of
states needed to expand the ground-state wave function
of the superblock grows exponentially

with the interblock interaction.
Since the density of states doubles

when the block size is doubled,
keeping a constant number of states will

not be enough unless the interactions
between two blocks also decrease exponentially.
For the usual lattice Hamiltonian
that we are interested in, the interblock interactions are fixed and
the conventional renormalization-group method breaks down.

In contrast to this behavior, explicit

examination of the matrices for

one-dimensional systems
shows that for the eigenvectors selected from the density matrix
the off-diagonal interblock interaction matrix elements decay
exponentially with the diagonal energy

separation. The interblock matrix

elements are only large between a few states.
It is this property that makes the DMRG method highly accurate.

In order for the total energy to be accurately calculated,
the block energies must be accurate. In the DMRG method,
the intrablock Hamiltonian matrix is
not in diagonal form. However, explicit
examination of the Hamiltonian shows that the off-diagonal
matrix elements are only large between a

few states close in diagonal
energy. Such a matrix with $|H_{ij}|

<< |H_{ii}-H_{jj}|$ at large $|i-j|$
describes a particle going uphill
in a one-dimensional potential ($H_{ii}$)

and with short-range hopping. We
know that in this problem the ground-state wave function
is exponentially small in the classically

forbidden region (at large $i$).

States deep inside the classically forbidden

region can be discarded.
Therefore, the low-energy states of the block

in the DMRG method are well

represented.

The extended states are good for the block

energy and the localized states
near the boundary are good for interblock matrix elements.
Thus, the selection of states in the DMRG

method is a compromise between
describing the block energy well and describing the interaction
between blocks well. This arises the interesting possibility
of satisfying both constraints without knowing the block
density matrix from the ground-state wave function.

We now prove that the ground-state energy

calculated in the finite-cluster
DMRG method is always a variational upper bound to the
true ground-state energy.
Our proof is based on the variational principle and the
facts that (i) the ground-state energy
calculated in the DMRG method is an eigenvalue

of the Hamiltonian in a truncated
Hilbert space, the dimension of which is

smaller than the dimension of the full
Hilbert space; (ii) the Hamiltonian

matrix elements calculated are

unaffected by the truncation of states

at each steps, {\it i.e.}, the
Hamiltonian matrix would have been the

same if all states were used in the
calculation. This is because the

truncated states can all be written
as linear combinations of the original

basis on which $c_i$ acts through
a series of orthogonal transformations.

Aside from the numerical errors,
the matrix elements are exact.

We would like to point out also that the sum of the weights (the

eigenvalues of the density matrix)
of states retained does not

necessarily indicate how close the
variational trial state is to the

true ground state. For example,
in going from an $l$-site block to

an $(l+1)$-site block,
the wave function is first obtained in

the Hilbert space of $m$ states from

the $l$-site block and two states of the

extra site. Then the space is
truncated to $m$ states in the $(l+1)$-site block.
The sum of the weights
measures the weight of the retained states relative
to the approximated ground-state
wave function in the enlarged Hilbert space of

dimension $2m$ and does not
necessarily correspond to the weight relative

to the true ground state.

In conclusion, the DMRG method is a

powerful self-consistent method for

finding the approximate ground state of a Hamiltonian.  It

automatically searches for the fixed

dimensional subspace that provides

the best expansion for the ground-state

wave function. At each iteration,
the states in the subspace are re-expanded

in terms of the original Hilbert
space of the states of a site, allowing

a rapid improvement in the

self-consistent iterations. The ground-state

energy is calculated variationally
in the best subspace found. For an $L \times L$ lattice,
the dimensions of the subspace needed for accurate
calculation grow exponentially with the linear

system size $L$.
However, for coupled chains ($L_x \times L_y$

with $L_y\ge 2L_x$), the
dimensions of the subspace become independent

of the chain length $L_y$

and depend only on $L_x$. Finally, our data suggest that
the number of states needed depends only on the total interaction

strenght between two blocks and is almost

independent of the number of

boundary sites.

We benefited from discussions with Clare Yu, Steve

White, and James F. Annett.
The work is supported in part by the Office of Naval Research
under Grant No. N00014-92-J-1340.
S. L. appreciates the hospitality of the Institute for
Theoretical Physics at University of California,
Santa Barbara where
the research was supported in part by NSF under Grant No.
PHY89-04035.

\begin{enumerate}

\item K. G. Wilson, Rev. Mod. Phys. {\bf 47}, 773 (1975).
\item J. W. Bray and S. T. Chui, Phys. Rev. {\bf B 19}, 4876 (1979).
\item J. E. Hirsh, Phys. Rev. {\bf B 22}, 5259 (1980).

\item S. R. White and R. M. Noack, Phys. Rev. Lett. {\bf 68}, 3487
(1992).
\item S. R. White, Phys. Rev. Lett. {\bf 69}, 2863 (1992);

 and unpublished.
\item P. W. Anderson, Science {\bf 235}, 1196 (1987); {\bf 256}, 1526
(1992);
{\bf 258}, 672 (1992).
\item The needed computation is roughly proportional to $m^3$.
\item For example, with $m=10$ we compute the

two-dimensional energy $E(m=10)$

starting from a one-dimensional lattice and then

gradually increase $m$ to $80$

where the calculated energy will be

very close to the exact energy. If we

now use the $m=80$ solution as the

environment block and reduce $m$ to 10
again, the energy obtained will be the same as $E(m=10)$ calculated

previously.
\item B. N. Parlett and D. S. Scott, Math. Comp. {\bf 33}, 217
(1979).
\item For this reason, it is difficult to impose the periodic
boundary

conditions in the $y$ direction. Presumably

this difficulty can be overcome
by using a periodic one-dimensional lattice to

prepare the environment blocks.

\item For free electrons, the decay rate $m^*$

will be the square of the

spinless case because spins up and down are

independent. The one-band Hubbard
model should require somewhat fewer states

than the free-electron case because
the reduced weight of double-occupied sites

effectively reduces the number of
states per site.  But on the other hand, when

the low-energy density of states
is high, the calculation will be more difficult.
\item D. J. Thouless, Phys. Rev. Lett. {\bf 39}, 1167 (1977).

\end{enumerate}

FIG. 1.  A square lattice is mapped to a one-dimensional lattice
along the solid line. There are two kinds of nearest-neighbor bonds
indicated by the solid lines with

$t_{ij}=1$ and the dashed lines with $t_{ij}=t'$.

FIG. 2.  (a) Representative data on how the

relative error in energy calculated
by the DMRG method $\Delta E/|E|$ depends

on the number of states kept for
the block $m$ for the $6\times 6$ lattice

with $t'=1$ at half filling.

(b) Number of states needed [$m^*$ in Eq.

(2)] as a function of the

system size $L$ for $L\times L$ lattices

with $t'=1$ at half filling.

Values of $1/m^*$ are obtained from the

slopes of semilogarithmic plots such
as (a).  Errors are about a few percent.

All the calculations were done on
workstations.

\end{document}